\documentclass{PoS}

\usepackage{graphicx}
\usepackage[figuresright]{rotating}

\newcommand{\AmS}{{\protect\the\textfont2
  A\kern-.1667em\lower.5ex\hbox{M}\kern-.125emS}}
\newcommand{\infl}{\chi} 

\newcommand{\mh}{m_{_{\rm H}}}
\newcommand{\mw}{m_{_{\rm W}}}
\newcommand{\gw}{g_{_{\rm W}}}
\newcommand{\gy}{g_{_{\rm Y}}}

\newcommand{\be}{\begin{equation}}
\newcommand{\ee}{\end{equation}}
\newcommand{\ba}{\begin{array}}
\newcommand{\ea}{\end{array}}
\newcommand{\baa}{\begin{array}}
\newcommand{\eaa}{\end{array}}
\newcommand{\bea}{\begin{eqnarray}}
\newcommand{\eea}{\end{eqnarray}}
\newcommand{\half}{{1\over2}}
\newcommand{\trace}{{\rm Tr}}

\newcommand{\Y}{{\rm Y}}
\newcommand{\Z}{{\rm Z}}
\newcommand{\EM}{{\rm em}}

\PoS{PoS(LAT2005)242}

\title{Magnetic field production after inflation. }

\ShortTitle{Magnetic field production after inflation. }

\author{\speaker{Andr\'es D\'{\i}az-Gil}\thanks{Supported by a FPU grant of the MECD.}\\
        Departamento de F\'{\i}sica Te\'orica\\
        Universidad Aut\'onoma de Madrid, Cantoblanco, 28048 Madrid, Spain\\
        E-mail: \email{andres.diazgil@uam.es}}

\author{J. Garc\'{\i}a-Bellido\\
        Departamento de F\'{\i}sica Te\'orica\\
        Universidad Aut\'onoma de Madrid, Cantoblanco, 28048 Madrid, Spain\\
        E-mail: \email{juan.garciabellido@cern.ch}}
\author{
        M. Garc\'{\i}a P\'erez \thanks{Supported by a Ram\'on y Cajal contract of the MECD}\\
Instituto de F\'{\i}sica Te\'orica UAM/CSIC\\
        Universidad Aut\'onoma de Madrid, Cantoblanco, 28048 Madrid, Spain\\
        E-mail: \email{margarita.garcia@uam.es}}
\author{
        A. Gonz\'alez-Arroyo\\
        Departamento de F\'{\i}sica Te\'orica\\
        Universidad Aut\'onoma de Madrid, Cantoblanco, 28048 Madrid, Spain\\
        E-mail: \email{antonio.gonzalez-arroyo@uam.es}}

\abstract{We study the electromagnetic field production during preheating 
after hybrid inflation in a model
with the field content of the Standard Model, coupled to a singlet inflaton.
We find that very soon after symmetry breaking our system
enters a regime of kinetic turbulence, characterized by a
self-similar behaviour of the energy spectra and a power-like
dependence on time of the inflaton and Higgs field variances. 
}

\FullConference{XXIIIrd International Symposium on Lattice Field Theory\\
                 25-30 July 2005\\
                 Trinity College, Dublin, Ireland}

\begin{document}

\section{INTRODUCTION}

There is a large amount of indirect observations of the presence of large scale 
cosmological magnetic fields (LSMF) with scales that go from the size of 
galaxies to the size of superclusters, with a relatively big
strength~\cite{magnetic}. 
It is believed that these magnetic fields were present in the form of a 
magnetic seed before the epoch of structure formation and that
they were amplified to its present magnitude.
Although there are several attempts to explain these magnetic field
seeds, their origin is still a mystery~\cite{magnetic}. It is tempting to 
speculate whether
they could be generated within the inflationary paradigm~\cite{infla}.
Our present work is along these lines. Taking the end of inflation
at the electroweak scale, we propose to place the origin of LSMF in the very
out of equilibrium epoch called the preheating era.
Our first aim is to throw some light into the process of electromagnetic 
field production during this period.

\subsection{The model and the method}

We consider a  hybrid inflation model which has the field content of 
the Standard Model (SM) gauge-scalar sector plus 
a singlet scalar field, the inflaton,  coupled to the SM Higgs:
\be
{\cal L} =- {1\over4}G^a_{\mu\nu}G^{\mu\nu}_a - {1\over4}F^\Y_{\mu\nu}F_\Y^{\mu\nu}+
  {\rm Tr}\Big [(D_\mu\Phi)^\dag D^\mu\Phi\Big ]
+ \half (\partial_\mu\infl)^2 - V(\Phi,\infl) \,,
\ee
with $\Phi=\half(\phi_0\,1\!{\rm l}+i\phi^a\tau_a)$ the Higgs field, 
$\infl$ the inflaton and 
$G^a_{\mu\nu}=\partial_\mu A_\nu^a - \partial_\nu A_\mu^a +
\gw \epsilon^{abc} A_\mu^b A_\nu^c$ and 
$F_{\mu\nu}^\Y=\partial_\mu B_\nu- \partial_\nu B_\mu$ the field strengths 
of the SU(2) and U(1) gauge fields respectively. The covariant derivative is:
$D_\mu = \partial_\mu - {i\over2}
\gw A_\mu^a\tau_a - {i\over2} \gy B_\mu$, with $\gw$ the SU(2) gauge coupling
and $\gy$ the hypercharge U(1) coupling.
The scalar potential includes a coupling to a massive inflaton:
\be
{\rm V} (\Phi,\infl) = 
{\rm V}_0 + \half(g^2\infl^2-m^2)\,|\phi|^2 + \frac{\lambda}{4}
|\phi|^4 + \half \mu^2 \infl^2 \,,
\ee
where $|\phi|^2\!\equiv\!2{\rm Tr}\, \Phi^\dag\Phi$, 
$\mu$ is the inflaton mass in the false vacuum 
and $m\!\equiv\!\sqrt\lambda\,v$; $\,v\!=\!246$ GeV.
During the period of hybrid inflation the inflaton field is dominated 
by its homogeneous mode $\chi_0$. Inflation ends at $t\!=\!t_c$ when this mode 
slow-rolls below $\chi_c\!\equiv\! m/g$ where the Higgs becomes massless.
Around this time  $\chi_0 \!=\! \chi_c (1-Vm (t-t_c))$,
with $V$ the inflaton dimensionless velocity. 
After $t_c$ a negative time-dependent mass-squared of the Higgs
is induced.  During the subsequent 
evolution, the low momentum modes of the Higgs grow exponentially 
in a process known as ``tachyonic preheating''~\cite{GBKLT}.
Preheating in this model has been proposed in
Refs.~\cite{Garcia-Bellido,Trodden}
as a mechanism for electroweak baryogenesis, an appealing possibility that has
been numerically investigated in \cite{Garcia-Bellido,jmt,smit0}.

To compute the time evolution of the system after inflation ends, we have
used the classical approximation (details can be found in~\cite{jmt} - 
see also~\cite{smit}).
For suitable parameter choices, the growth of the Higgs field low momentum 
modes takes place before non-linearities in the Higgs potential and 
the coupling to the gauge fields have become relevant. 
Thus, one can exactly  compute the quantum evolution of the Higgs
field. It is such that, after 
a while, the tachyonic modes of the Higgs evolve like classical 
modes. This initial stage is very fast and therefore the remaining  degrees
of freedom can be assumed to remain in their initial quantum vacuum (ground) 
state. Once sufficient number of Higgs modes have become classical
we substitute the exact quantum evolution by the classical one. For 
consistency, this substitution should be done before non-linearities have
become relevant.
The advantage of this approximation is that it allows to
treat the full non-linear evolution of the system by lattice techniques.

\section{RESULTS}

The results presented here correspond to the particular choice of parameters
denoted as Model A1 in Ref.~\cite{jmt}. For this model 
the inflaton velocity $V\!=\!0.024$,  
$g^2\!= \!2 \lambda$, $\gw\! =\! 0.05$, and the Higgs to W boson mass ratio is given by 
$\mh/\mw\!=\!4.65$. Here we also introduce the U(1)
hypercharge field, fixing $\gy$ to reproduce the 
experimentally determined Weinberg angle. This choice of parameters,
although far from the physical value of $\mw$, allows an optimal control
of lattice and finite volume artefacts for the lattices sizes 
reachable with our present computer resources~\cite{jmt}. Results for physically
relevant parameters will be presented elsewhere~\cite{ajmt}. 
The chosen values of the lattice cut-off are $ma\!=\!1.31$ (NS=32), 
$ma\!=\!0.87$ (NS=48) and $ma\!=\!0.655$ (NS=64). 
The time-like lattice spacing $a_t$ must be smaller than
the spatial one for the stability of the discretized equations of motion.
We have chosen $a /a_t  = 40$ and $80$ and checked stability of the 
results under this change.

\vspace*{-0.2cm}
\subsection{Means and Spectra}

\begin{figure}[htb]
\vspace{4.8cm}
\includegraphics{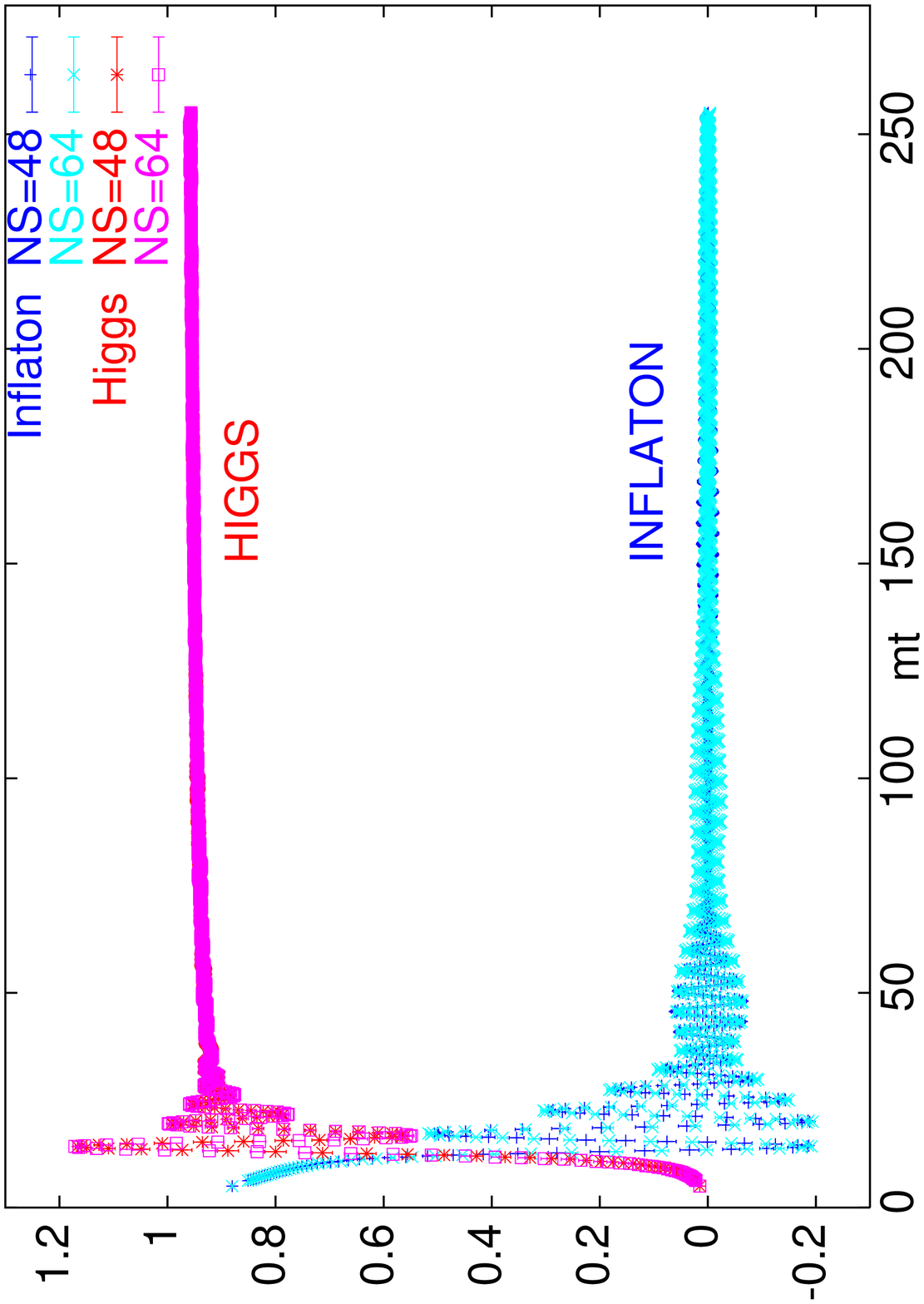}
\includegraphics{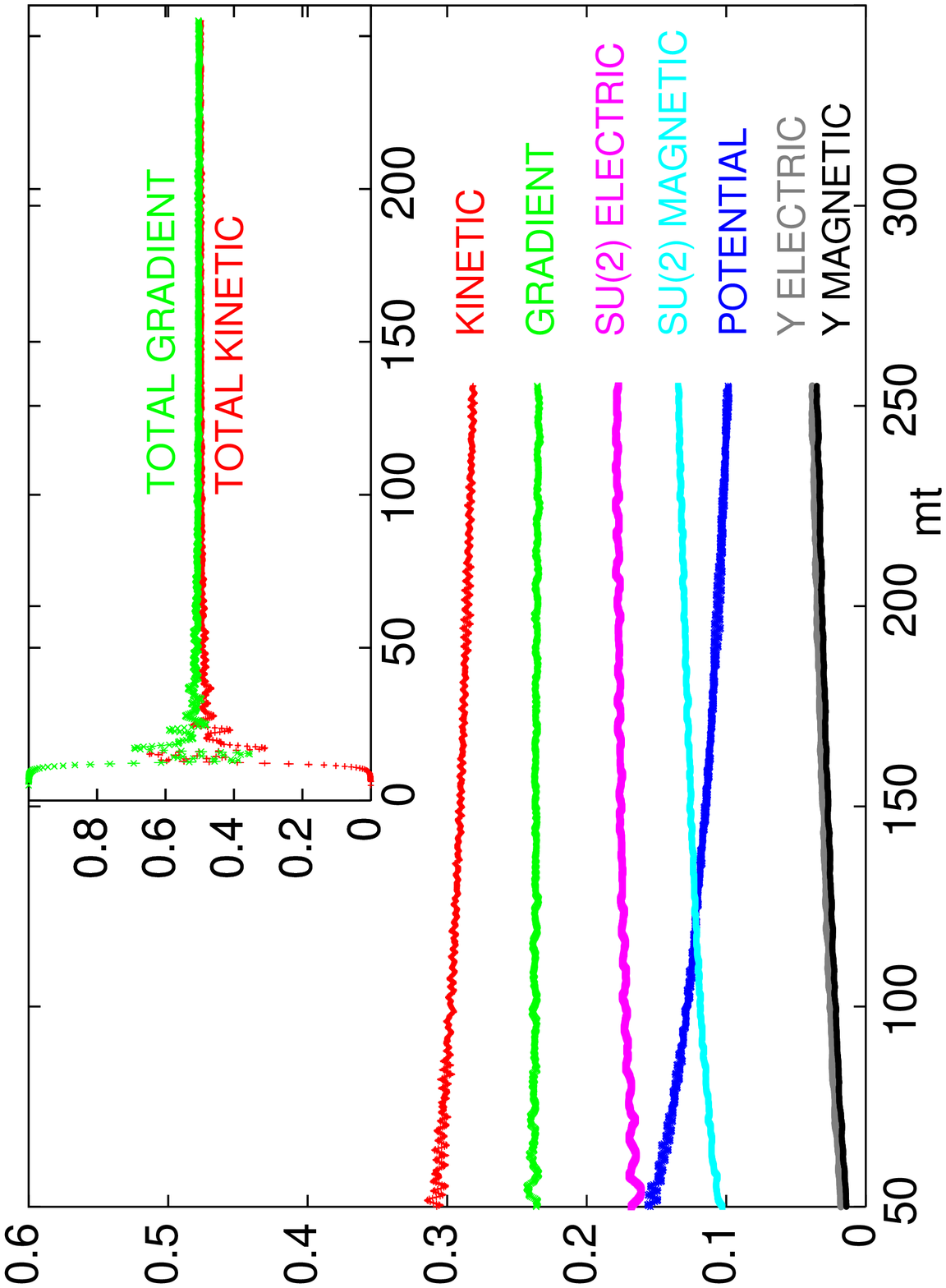}
\caption{Left: The time evolution of $\langle \overline{|\phi|}/v\rangle$,
$\langle \overline{\chi}/\chi_c\rangle$. Right: Late time behaviour of the 
different components of the energy, normalized to the total energy, for NS=48.}
\label{fig1}
\end{figure}

In Fig.~\ref{fig1} (Left), the time evolution of $\langle \overline{|\phi|}/v\rangle$ 
and $\langle \overline{\chi}/\chi_c\rangle$ illustrates 
the process of symmetry breaking: the Higgs starts in the false vacuum at 
$t=t_c$ and quickly evolves into the symmetry broken phase. 
Fig.~\ref{fig1} (Right) shows the late time behaviour of the different 
components of the energy, normalized to the total energy, for $ma\!=\!0.87$.
The evolution is such that kinetic and potential
energies virialize as for harmonic oscillators ($\langle$kinetic$\rangle \!=
\!\langle$potential$\rangle$). 

\begin{figure}[htb]
\vspace{5.4cm}
\includegraphics{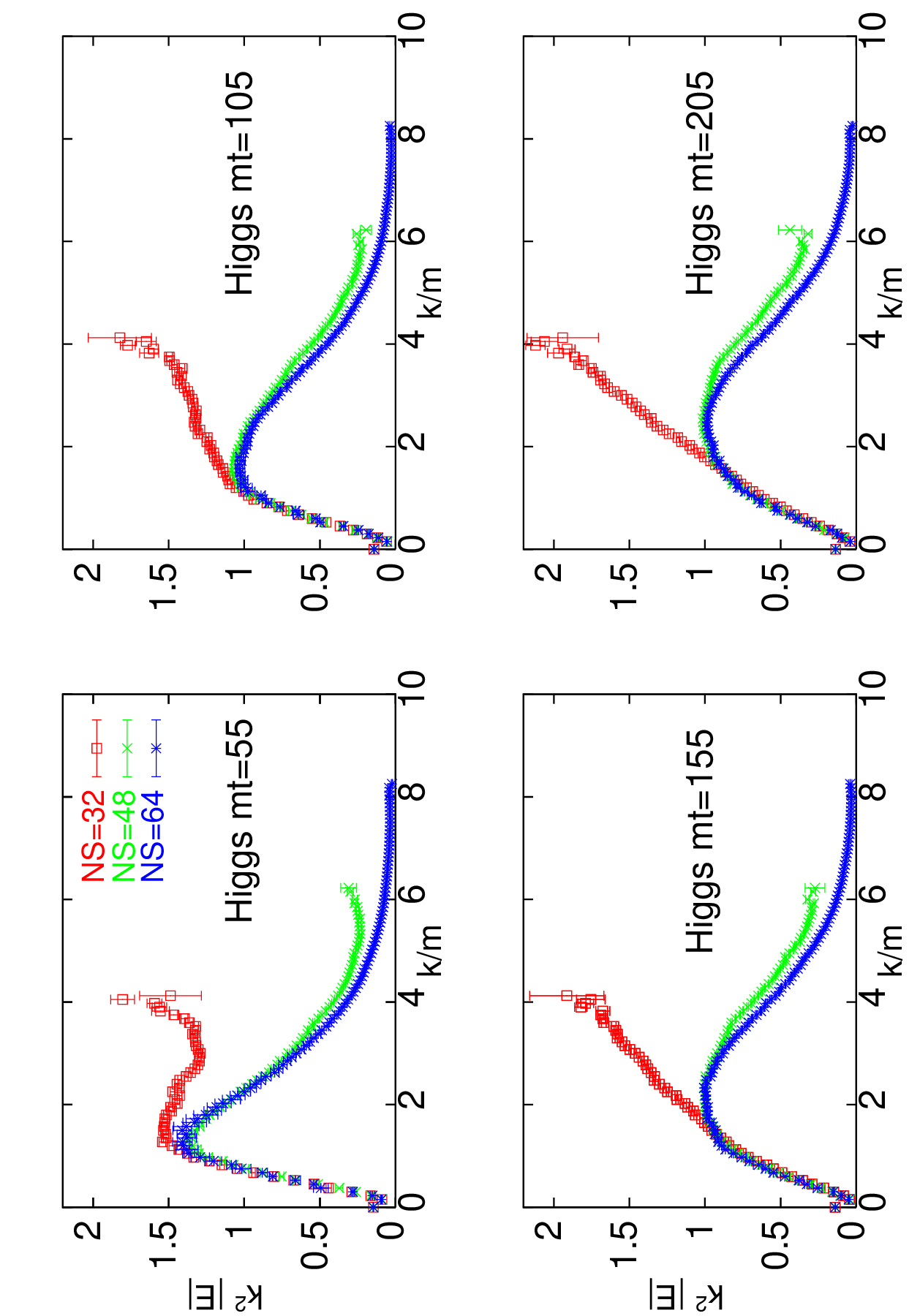}
\includegraphics{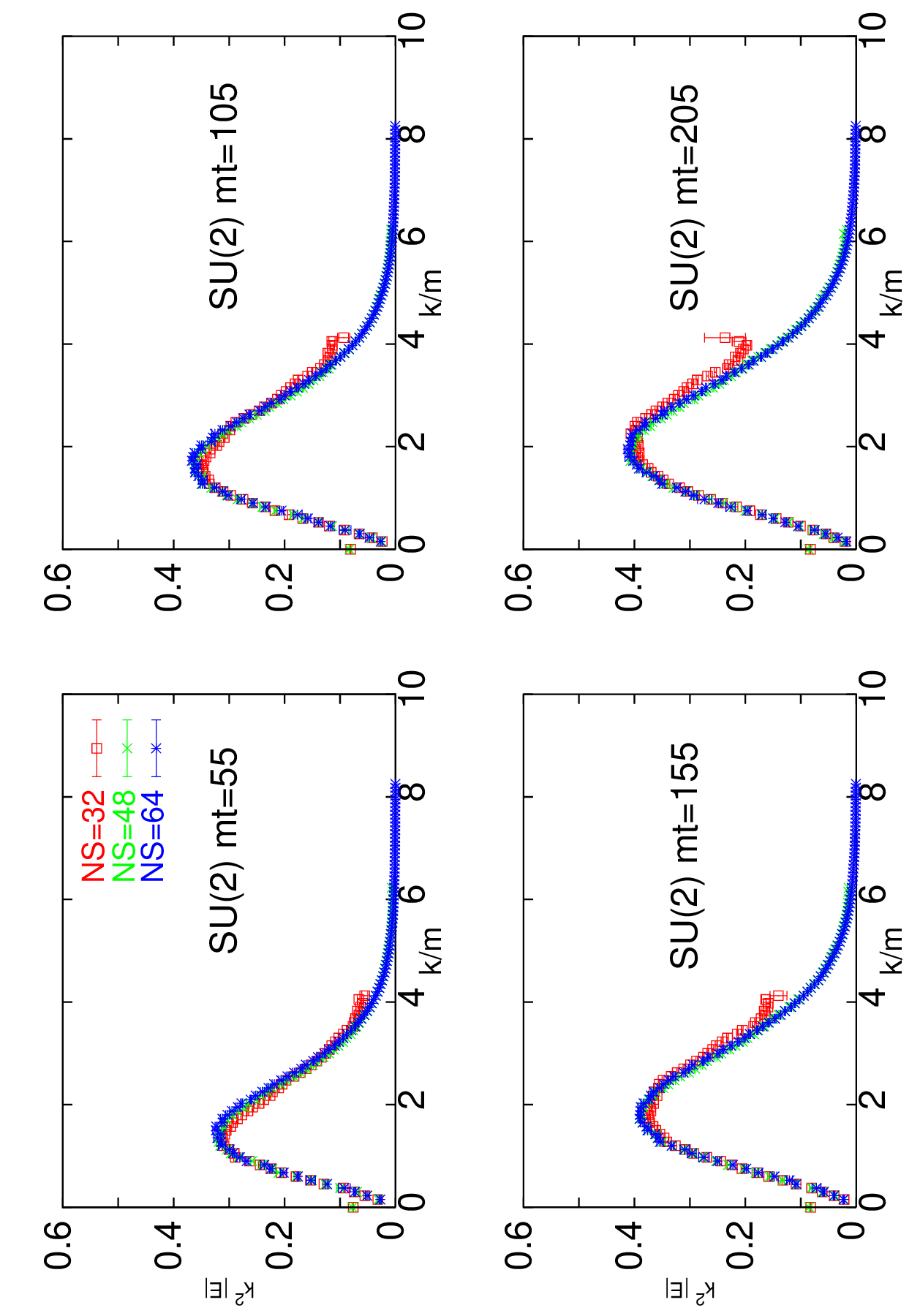}
\caption{Spectra of the inflaton+Higgs kinetic
energy and SU(2) electric energy. 
}
\label{spectra}
\end{figure}

We are interested in the behaviour of electromagnetic fields 
long after symmetry breaking. 
It is not obvious that our approximations remain under control for 
such late times. The validity of the classical
approximation rests on the fact that the tachyonic infrared modes dominate the
dynamics. This cannot happen if ultraviolet modes are strongly excited
when the physically relevant processes take place. 
A way to expose the range of relevant momenta is to analyze the Fourier 
spectra of the (gauge invariant) energy densities.
Fig.~\ref{spectra}  shows, as an illustration, 
the time evolution of the  
spectra of the inflaton+Higgs contribution to the kinetic energy
and of the electric component of the SU(2) gauge field energy.
To enhance the effect of lattice artefacts for large momenta,
we plot $k^2 |E(k)|$. Note that the presence of the lattice cut-off truncates
the range of allowed momenta. The number, $N(k)$, of lattice momenta of norm
$k$
should be proportional to $k^2$
in the range of relevant momenta for the cut-off not to distort the
physical spectrum. Deviations of $N(k)$ from the $k^2$ behaviour appear
for $k/m>$  2.5, 3.5, 4.5 for NS= 32, 48, 64 respectively. It is clear from
the figure that the NS=32 lattice is off.
NS=48 starts showing deviations, for physically relevant momenta, for times
above $mt \! \sim \! 100$. These deviations, however, seem to remain small up
to much larger times. Full checks of  cut-off independence by going to
larger lattices will be done in the near future.
Nevertheless, given these results, our approach to study the evolution
up to relatively large times, $mt\! \sim \! 250$, seems viable. The time ranges
in this study are four times larger than those in Ref.~\cite{jmt}.

\vspace*{-0.5cm}
\subsection{Electromagnetic field production}

In order to analyze the production of electromagnetic fields we have,
first, to extract the U(1)$_{\rm em}$ content of the SU(2)$\times$U(1) fields
in the Lagrangian. Details of the lattice implementation will be given 
elsewhere~\cite{ajmt}. In terms of the SU(2) link, $U_\mu(x)$, and the hypercharge
link, $\hat B_\mu(x)$, one can express the field associated to
the $Z$ boson as:
\be
\hat Z_\mu (n) = -\trace \Big ( i \tau_3 \ {\Phi^\dagger (n) \over |\phi(n)|}
\  U_\mu (n) \ {\Phi( n+\mu) \over |\phi(n+\mu)|} \ \hat B_\mu(n) \Big )
\hspace{5mm} \stackrel{a\rightarrow 0}{\longrightarrow} \hspace{5mm}
a_\mu g_Z Z_\mu \,, \label{zeta}
\ee
with $a_{\mu \ne 0} = a$, $a_0=a_t$ and $g_Z$ the $Z$ boson coupling.
The field strength of the U(1)$_{\rm em}$ field is obtained from:
$
\hat F^{\EM}_{\mu \nu}(n)^2  = 1-\cos \Big ({\gy^2 \over \gy^2 + \gw^2} 
\hat F^{\Z}_{\mu \nu}(n) - \hat  F^{\Y}_{\mu \nu}(n)\Big ) \,,
\label{fzeta}
$  
with $ F^\Z_{\mu \nu} (n) \!=\! \hat Z_\mu (n)\!-\!\hat
Z_\nu (n+\mu)\!-\!\hat Z_\mu (n+\nu)\!-\!\hat Z_\nu (n)$ and
$\hat F^\Y_{\mu \nu} (n) \!=\! \hat B_\mu (n)\!-\!\hat B_\nu (n+\mu)
\!-\!\hat B_\mu (n+\nu)\!-\!\hat B_\nu (n)$.
In the continuum limit $\hat F^{\EM}_{\mu \nu} \longrightarrow a_\mu a_\nu e 
F^{\EM}_{\mu \nu}$, with $F^{\EM}_{\mu \nu}$ the continuum electromagnetic 
field strength.

In computing the field strength of the $Z$ boson, $\hat F^\Z_{\mu \nu}$, from 
Eq.~(\ref{zeta})
there is a complication that arises due to the possible discontinuous 
behaviour of $\Phi^\dagger (n) / |\phi(n)|$  
at points where the Higgs field is zero. At such points the $Z$ boson and the  
electromagnetic field strengths can show a (lattice) delta-function behaviour. 
This gives rise to the appearance of non-scaling spikes
in the corresponding energy densities which affect also spatially averaged
quantities.  Although configuration averaging
smears out the effect of the spikes, they give rise
to a very noisy signal. One way to improve the signal 
is to eliminate from the spatial averages the points where the 
Higgs norm gets closer
to zero than a certain value $|\phi_c|$.  Taking $|\phi_c|=0.08$
considerably reduces the noise and  gives results compatible within errors
with those obtained from larger statistics sets.
The appearance of these spikes is rather common in the first stages of 
the evolution where the Higgs field gets rather often close to zero. It becomes
more rare as time evolves, although occasionally they can also occur at times
at large as $mt=200$. It is attractive to speculate that these zeroes
could be associated to sphaleron-like configurations, a point that deserves
further investigation~\cite{ajmt}.
 
\begin{figure}[htb]
\vspace{4.5cm}
\includegraphics{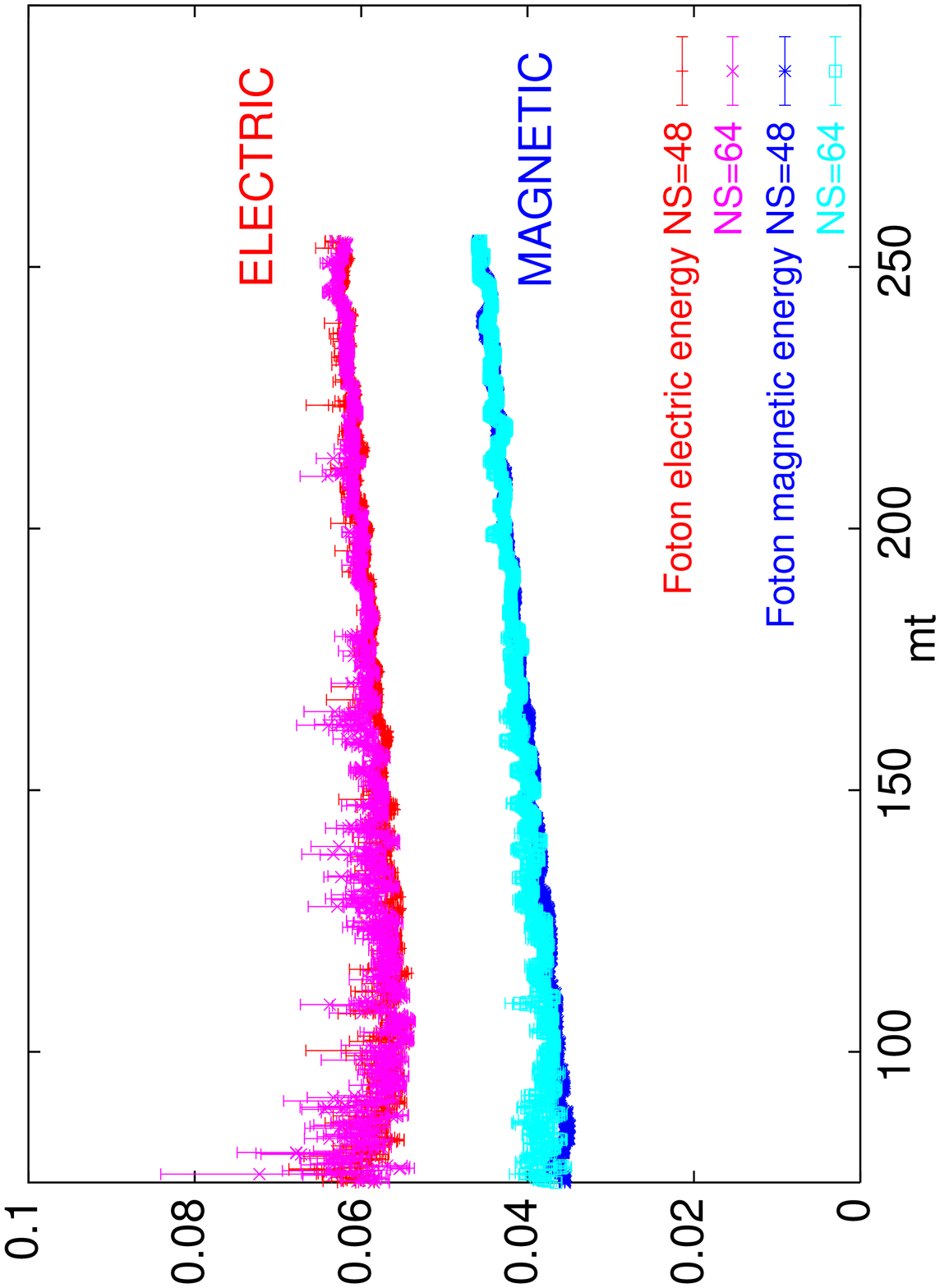}
\includegraphics{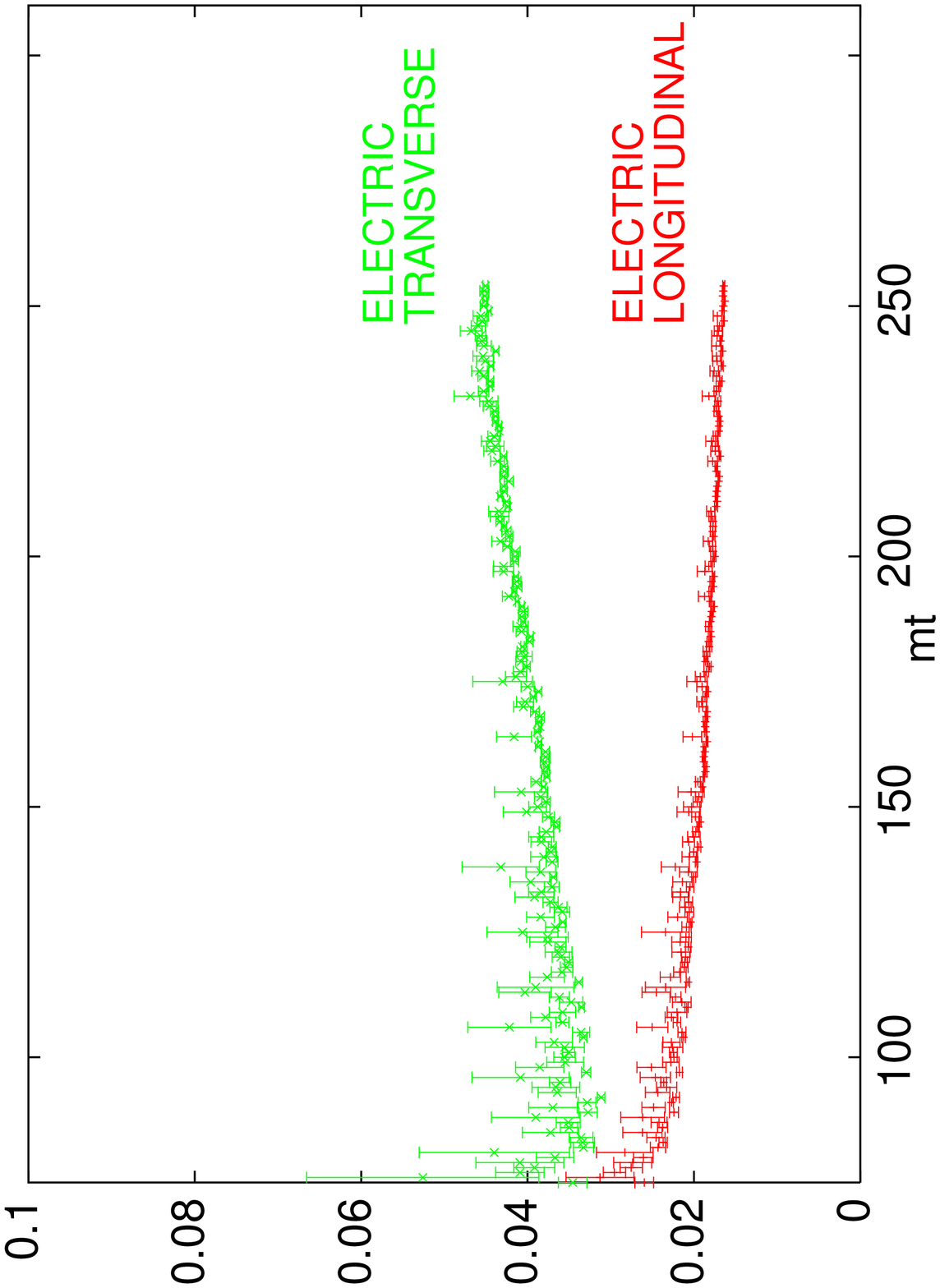}
\caption{Left: Late time  behaviour of the
electric and magnetic components of the U(1)$_{\rm em}$ energy, normalized to
the total energy. Right: Longitudinal and transverse parts of the electric component.} 
\label{ele3}
\end{figure}

Figure~\ref{ele3} shows our final result for the late time  behaviour of the
electric and magnetic components of the U(1)$_{\rm em}$ energy, normalized to
the total energy. Results are shown for two different lattice spacings,
agreeing well within errors.
A detailed analysis of the properties of the electromagnetic field will
be deferred to a future paper~\cite{ajmt}. Here we only want to point out that a
separation between the longitudinal and transverse components of the
electric field can be easily performed by going to Fourier space. The result
is shown in  Fig.~\ref{ele3} (Right). The longitudinal component
decreases with time. This is the expected behaviour at large times since 
the system evolves into a plasma with charge density going to zero.

\begin{figure}[htb]
\vspace{4.5cm}
\includegraphics{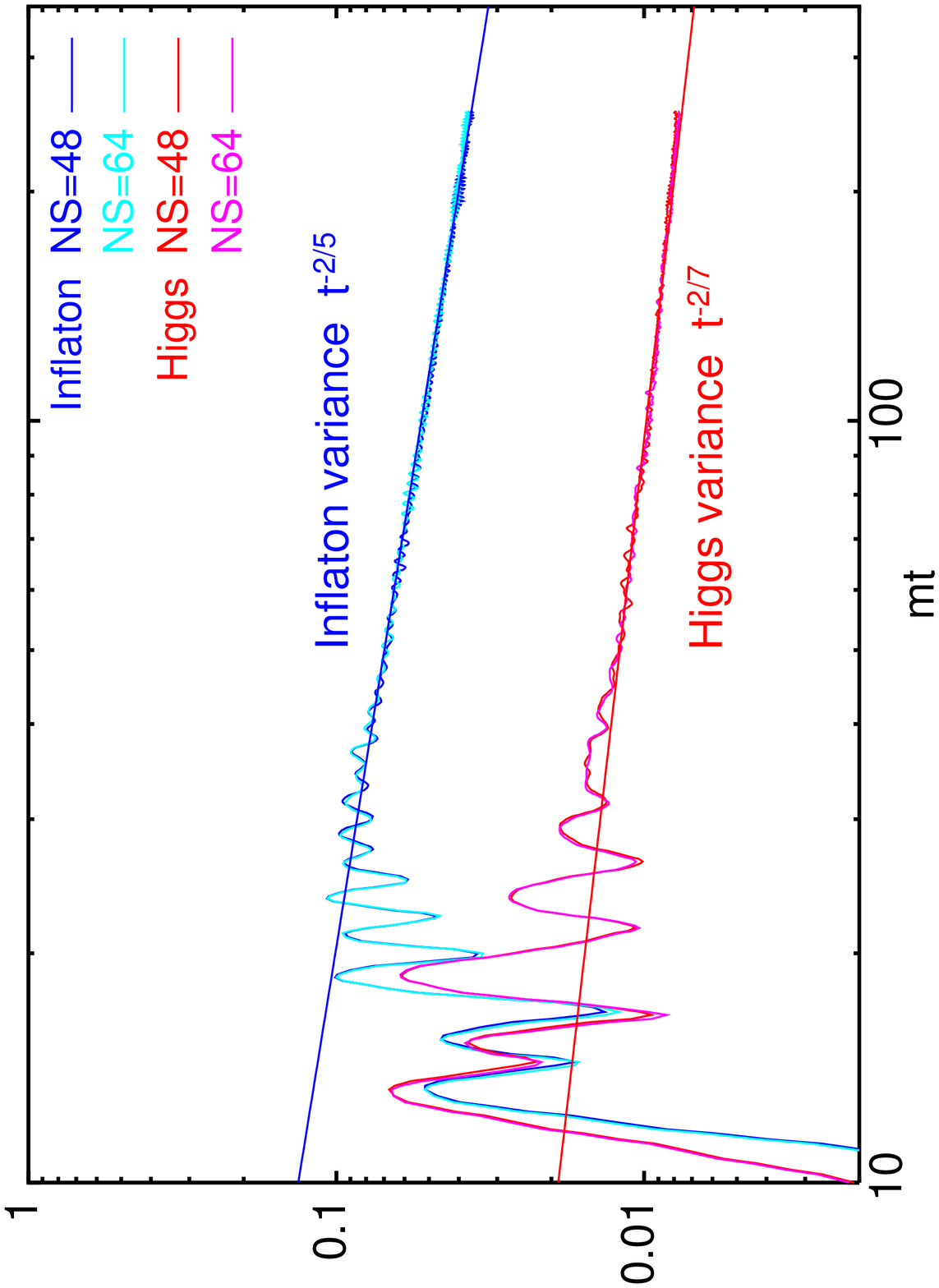}
\includegraphics{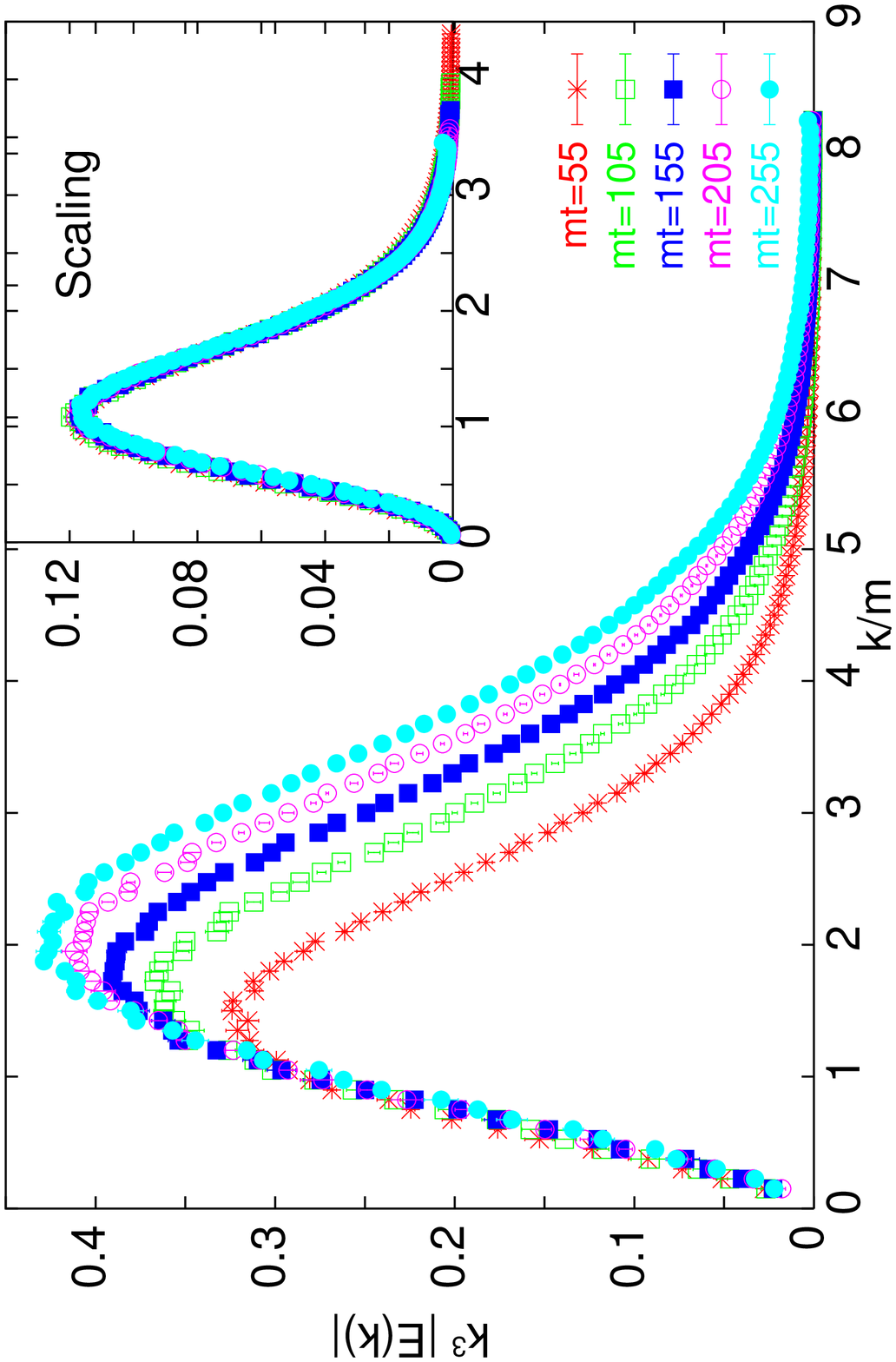}
\caption{Left: The late time behaviour of the variance of the Higgs and inflaton
fields. Right: Spectra of the SU(2) electric part of the energy
showing a self-similar scaling according to Eq.~(2.2).}
\label{kin1}
\end{figure}

\vspace*{-0.4cm}
\subsection{Kinetic turbulence}
According to Ref.~\cite{micha},
systems like the one we study evolve towards thermalization via
a period of kinetic turbulence.
This regime is characterized by a specific behaviour of the variance of
the scalar fields together with a particular scaling of the energy spectra.
To be concrete, one expects for the scalar fields:
$
\langle \phi^2\rangle - \langle \phi\rangle^2 \sim t^{-\nu} 
$
with $\nu = 2/(2m-1)$ in three spatial dimensions and for $m$-particle 
interactions.
At the same time the energy spectra show a self-similar behaviour:
\be
n(k,t) = t^{-q} n_0 (k t^{-p}), \hspace{1cm} {\rm with}
\hspace{5mm} n(k,t) \equiv {E(k,t) \over k}
\label{eq:kin2}
\ee
where $q=3.5p$ and $p=1/(2m-1)$. Ref.~\cite{micha}  analyzes models
containing only scalar fields. The characteristic times for the kinetic
turbulence regime are $mt > 3000$ after inflation ends. 
We also find that our data enter a regime of kinetic turbulence, 
but this happens at much earlier times $mt>50$, presumably due 
to the larger number of degrees of freedom.
The variances of the Higgs and inflaton fields are shown in 
Fig.~\ref{kin1} (Left).
They follow the expected behaviour  with $m\!=\!4$ for
the Higgs and $m\!=\!3$ for the inflaton. 
Fig.~\ref{kin1} (Right) shows the spectra of the electric part of 
the SU(2) energy
for several times from $mt=55$ up to $mt=250$. The data remarkably follow 
the self-similar behaviour of Eq.~(\ref{eq:kin2}) with $p=1.1/7$, very close to 
the value expected for $m\!=\!4$.
The early onset of kinetic turbulence is good news since it allows 
extrapolation of the time evolution beyond the limitations set by our 
numerical approach.

\section{CONCLUSIONS} 

Making use of the classical approximation, we have studied the production of 
electromagnetic fields during preheating after a period of low-scale
hybrid inflation. We present here a feasibility study showing how
our approach reliably describes the evolution up to rather large
times, $mt\!\sim\! 250$.

We have extracted the total electric and magnetic $U(1)_{\rm em}$ energies.
Concerning electric fields we observe a damping of the longitudinal 
component of the field strength, corresponding to the gradual
neutralization of charged particles in the primordial plasma.
The total magnetic field adds up the contribution of radiation and
the primordial magnetic seed. The separation of the latter
will be deferred to a future publication~\cite{ajmt}.
It is particularly interesting
the fact that, very soon after inflation ends ($mt> 50$), the SU(2) 
energy spectra show a self-similar behaviour
(see Fig.~(\ref{kin1})). This is a signature of kinetic turbulence.
 We still have to check that this seed gives rise to a
magnetic field correlated over large scales (comparable to the horizon), and
that it remains correlated for a sufficiently long time.


\begin{thebibliography}{9}
%
\bibitem{magnetic}
M.~Giovannini,
\emph{Int.\ J.\ Mod.\ Phys.}\ D {\bf 13} (2004) 391.
%
\bibitem{infla}
M.~Giovannini and M.~E.~Shaposhnikov,
\emph{Phys.\ Rev.\ D} {\bf 62} (2000) 103512.
%
\bibitem{GBKLT} G. Felder, J. Garc\'\i a-Bellido, P. Greene, L. Kofman,
A. D. Linde and I. Tkachev, \emph{Phys. Rev. Lett.} {\bf 87}, 011601 (2001);
G. Felder, L. Kofman and A. D. Linde, \emph{Phys. Rev. D} {\bf 64}, 123517 (2001
).
%
\bibitem{Garcia-Bellido}
J.~Garc\'{\i}a-Bellido, D.~Y.~Grigoriev, A.~Kusenko and M.~E.~Shaposhnikov,
\emph{Phys.\ Rev.\ D} {\bf 60} (1999) 123504.
%
\bibitem{Trodden}
L.~M.~Krauss and M.~Trodden,
\emph{Phys.\ Rev.\ Lett.}  {\bf 83} (1999) 1502.
%
\bibitem{jmt}
J.~Garc\'{\i}a-Bellido, M.~Garc\'{\i}a P\'erez and A.~Gonz\'alez-Arroyo,
\emph{Phys.\ Rev.\ D} {\bf 67} (2003) 103501;
\emph{Phys.\ Rev.\ D} {\bf 69} (2004) 023504.
%
\bibitem{smit0}
A.~Tranberg and J.~Smit,
\emph{JHEP} {\bf 0311} (2003) 016;
\emph{JHEP} {\bf 0212} (2002) 020.
J.~I.~Skullerud, J.~Smit and A.~Tranberg,
\emph{JHEP} {\bf 0308} (2003) 045.
%
\bibitem{smit}
J. Smit, \emph{Simulations in  Early-Universe Theory},
in these Procceedings and references therein.
%
\bibitem{ajmt}
A.~D\'{\i}az-Gil, J.~Garc\'{\i}a-Bellido, M.~Garc\'{\i}a P\'erez 
and A.~Gonz\'alez-Arroyo, in preparation.
%
\bibitem{micha}
R.~Micha and I.~I.~Tkachev,
\emph{Phys.\ Rev.\ D} {\bf 70} (2004) 043538.
R.~Micha and I.~I.~Tkachev,
\emph{Phys.\ Rev.\ Lett.}\  {\bf 90} (2003) 121301.


\end{thebibliography}
\end{document}